\documentclass[5p,times,authoryear]{elsarticle}

\usepackage{ecrc}

\volume{00}

\firstpage{1}

\journalname{Astronomy and Computing}

\runauth{K. Nolan, E. Hickey \& O. Creaner}

\jid{astron.comput.}

\jnltitlelogo{Astronomy and Computing}

\usepackage{amssymb}

\usepackage[figuresright]{rotating}

\usepackage[table,dvipsnames]{xcolor}

\usepackage{makecell}

\usepackage{mathtools}
\usepackage{caption}
\DeclareCaptionType{equ}[Expression][]

\begin{document}

\begin{frontmatter}



\dochead{}

\title{The Locus Algorithm II: A robust software system to maximise the quality of fields of view for Differential Photometry}

\author[ITTD]{Kevin Nolan}
\ead{kevin.nolan@tudublin.ie}

\author[ITTD]{Eugene Hickey}
\ead{eugene.hickey@tudublin.ie}

\author[ITTD,DIAS,LBNL]{Ois\'{i}n Creaner\corref{cor1}}
\cortext[cor1]{Corresponding author}
\ead{creanero@gmail.com, oocreaner@lbl.gov}

\address[ITTD]{Technological University Dublin, Tallaght Campus, Dublin 24, Ireland}
\address[DIAS]{Dublin Institute for Advanced Studies, 31 Fitzwilliam Place, Dublin 2, Ireland}
\address[LBNL]{Lawrence Berkeley National Laboratory, 1 Cyclotron Road, Berkeley, California, USA}

\begin{abstract}
We present the software system developed to implement the Locus Algorithm, a novel algorithm designed to maximise the performance of differential photometry systems by optimising the number and quality of reference stars in the Field of View with the target.  Firstly, we state the design requirements, constraints and ambitions for the software system required to implement this algorithm. Then, a detailed software design is presented for the system in operation. Next, the data design including file structures used and the data environment required for the system are defined.  Finally, we conclude by illustrating the scaling requirements which mandate a high-performance computing implementation of this system, which is discussed in the other papers in this series.

\end{abstract}


\end{frontmatter}


\section{Introduction}
\label{Introduction}
The Locus Algorithm, first proposed by \citet{creaner2010large} and defined in full in \citet{locuspaper} is an algorithm that adjusts the position of the Field of View (FoV) to provide optimised conditions for differential photometry, given a target and telescope parameters. It works by translating the FoV on a North-South/East-West basis such that the maximum number of highest quality of reference stars are included in the FoV and the target remains in the FoV.  A software system which harnesses this algorithm was developed as shown in \citet{creaner2016thesis} and detailed further here.  

This paper is part of a series on the Locus Algorithm, from its principles to its development and implementation and performance metrics of a grid implementation.  This paper focuses on the software required to implement the algorithm on single targets or on small batches of targets, which can be run on a single computer.  The system described here assumes that all required data can be made available on the device that is being used, and that outputs can be stored persistently on that device for the course of the project.  These assumptions are valid for testing purposes but do not remain so for the purposes of the creation of catalogues of pointings for large collections of target objects.

This system was used to develop two catalogues of pointings based on the Sloan Digital Sky Survey (SDSS) catalogue. The first catalogue started with 40,000 quasars from the SDSS 4th quasar Catalogue \citep{abazajian2009seventh} as input targets to generate pointings for 23,779 quasars as discussed in \citet{quasarpaper} and with 357,175,411 point sources from SDSS DR7 \citep{schneider2007sloan} as inputs to produce pointings for 61,662,376 stars for use as candidates for exoplanet observation as shown in \citet{ZenodoXOPCatalogue}. 

Creating catalogues of this scale is not practical with a single computer.  Instead, a High-Performance Computing (HPC) system is required to generate large catalogues of pointings in a practical timeframe.  To this end, the third paper in this series, \citet{grid_system_paper} discusses a grid computing solution was developed to enable the system to be parallelised and reach the required scale. The final paper of this series, \citet{grid_metrics_paper} explains that, with the HPC system developed and implemented, it was necessary to apply a series of metrics to it in operation and assess areas in which the system performed well, and where there was need for improvement.

\section{Design Considerations}
\label{Considerations}

There were a number of factors impacting the design of the software system, which are grouped into three categories as outlined below:  

Design \textbf{Requirements}: the elements of the project necessary for it to achieve a ``minimum viable product/output'' of this project.  

Design \textbf{Constraints}: Limiting factors in achieving the software design requirements such as source data formats.

Design \textbf{Ambitions}: Elements of the design which are necessary for it to perform beyond a ``minimum viable product'' and allow future project uses to be achieved.


\subsection{Requirements}
\label{requirements}

The core of this analysis process may be summarised in four central requirements. The software must: 

\begin{itemize}
\item Read data in from a source catalogue.
\item Identify potential pointings from within that data for a set of targets.
\item Compare those pointings to determine the optimal pointing for each target.
\item Output each target and its pointing.  
\end{itemize} 

Reading data from the source catalogue requires the selection of a suitable source catalogue for the task at hand, identifying its format, extracting the data from that format into memory and making it available for subsequent stages of the project.  For this project, the SDSS Calibrated Object list (\texttt{tsObj*.fit}) files available from the Data Archive Server (DAS) \citep{SDSSDAS} were selected because, during the planning phases of this project, SDSS represented the gold standard of optical astronomical catalogues \citep{jordi2006empirical}.  These files are in the Flexible Image Transport System (FITS) file format, the adoption of which influences the design of the software system as discussed in Subsection \ref{constraints}.

\begin{figure}[!htb]
        \center{\includegraphics[width=0.47\textwidth]
        {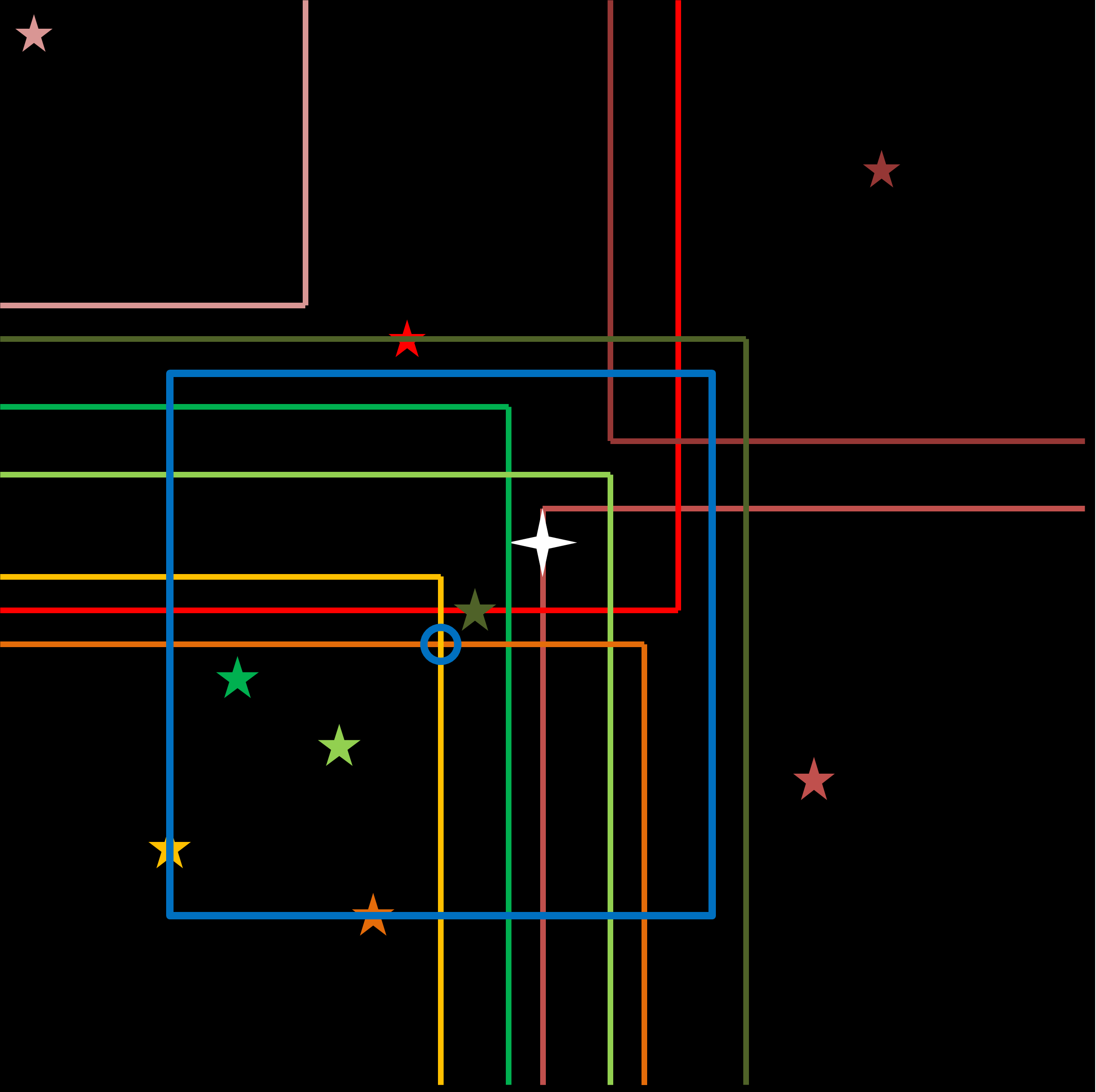}}
        \caption{\label{fig:locus} An example of an optimum pointing generated using the locus algorithm.  Modified from \citet{creaner2016thesis}}
      \end{figure}

The Locus Algorithm defines how pointings are identified and compared with one another.  It is detailed in \citet{locuspaper}, but summarised in the following nine steps, the results of which are illustrated in Figure \ref{fig:locus}.  

\begin{enumerate}
\item Identify the target by RA/Dec coordinates and its magnitudes in the system used (e.g. SDSS \textit{ugriz}).
\item Provide observational parameters: FoV size, magnitude and colour difference limits ($\Delta$mag\textsubscript{max} and $\Delta$col\textsubscript{max}) and resolution.
\item Define a \textit{Candidate Zone}(CZ) to identify all stars which could be included in a FoV with the target.
\item Filter the CZ to select \textit{Candidate Reference Stars} (CRS) based on the observational parameters.
\item Calculate a \textit{Rating} for each CRS based on how closely its colour matches that of the target.
\item Calculate Loci around each CRS upon which the FoV may be centred and include the CRS and the Target.
\item Identify Points of Intersection (PoI) between these Loci.
\item Calculate a \textit{Score} for each PoI by combining ratings.
\item Output the Optimum pointing which is the PoI with the highest Score.
\end{enumerate}

Each identified pointing is presented as a row in an output file containing the relevant information useful to make a optimized differential-photometric observation of the associated target:

\begin{enumerate}
\item The RA, Dec and magnitude of the target
\item The RA, Dec and Score associated with it's optimum pointing
\end{enumerate} 

The Identification of all relevant reference stars in the optimum field can also be subsequently sourced from the SDSS using this information.  To maintain consistency with input data, this output is also stored in FITS format.
 
\subsection{Constraints}
\label{constraints}

The SDSS data used in this system is supplied in FITS format.  The use of SDSS data in this system brought a series of constraints to the software
system developed. At the time the software was developed, the 'C' FITS library, CFITSIO, \citet{pence1999cfitsio} was the primary software tool available to access FITS files in
software. This constrained the development of the system in C; and since C is a
procedural language, this mandated a procedural approach to the development of the
system.

A final important requirement of the system worth identifying here is the necessity to
pre-process the SDSS input FITS files so as to filter out unnecessary components of
the input data, as well as to store the necessary input data components in a generic
format optimised for pipeline processing. Such functionality also links to a project
ambition discussed in Section \ref{API} of developing the capability of processing data from
various input data sources and not just from SDSS. Hence these combined
requirement give rise to the necessity of an input data Application Programming
Interface or API capable of addressing such input data requirements in both the
current version and in future revisions of the system.

Mindful of the scalability requirements to process huge catalogues of targets, early
design phase processing requirement estimates pointed to the necessity for the
completed system to run on a High Performance Computing System. While this paper
concentrates on the aspect of the system to analyze a single target, the HPC enabled
software system has also been completed and is described in full in\citet{grid_system_paper}.We provide here a summary of that HPC system components.

The HPC solution chosen was a Grid Computing solution, specifically, the National Grid Infrastructure (NGI) operated by Grid Ireland.  This system uses a \textit{glite} middleware architecture to distribute grid jobs among Worker Nodes (WN).  These WN are commercial-off-the-shelf computers each running separate instances of the Scientific UNIX operating operating system \citep{coghlan2005grid}.  As a result, the system developed here was constrained to run in a UNIX environment.  Beyond the operating system requirement, however, the grid middleware was designed to act as a ``wrapper'' around the core software discussed in this paper.  Grid management scripts were written to create environments such that the design of the software presented here was not constrained by the grid, following a principle of modularity - each element of the software could be developed in isolation.  Further details of this design constraint, and how it was mitigated are given in \citet{grid_system_paper}.

\subsection{Ambitions}
\label{ambitions}

In addition to meeting the requirements of this project, the design of the software was impacted by the requirement to be able to respond to future project ambitions, such as the ability to operate on different input catalogues and input formats of catalogues; the ability to modify or change core functionality of the software, and for it to be able to deliver outputs appropriate to those required changes. 

These additional requirements were catered for, as best as possible, in a software design that was flexible, modular, extensible and which enable data abstraction.

\textbf{Flexibility} enabled though maximum possible parameterization of all inputs to allow user-level input to effect significant changes in output.

\textbf{Modularity} All components to the system designed in a modular way to enable the future modification or replacement of future required aspects of the system.

\textbf{Extensibility} The system designed in a pipe-line fashion such that future additions to the system could be implemented with minimal impact to the existing system.

\textbf{Abstraction} in the data was implemented by creating a data format that could be used with any data source and extracting the data from the source into that format.  In doing so, data that would not be used in the algorithm was not copied to the new format, and data volumes could be reduced substantially by comparison to the original data

\section{Software Design}
\label{Design}

\begin{figure}[!htb]
        \center{\includegraphics[width=0.47\textwidth]
        {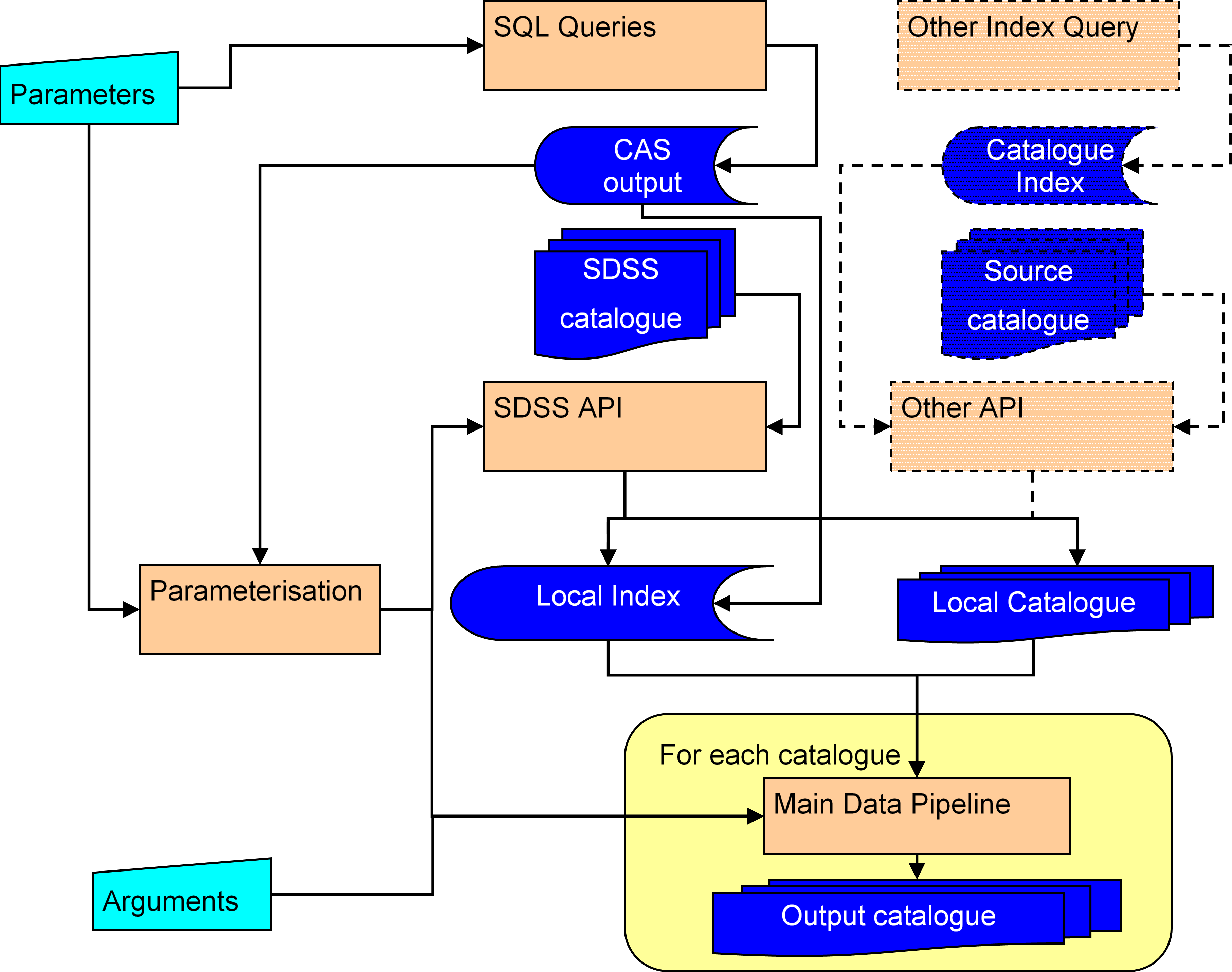}}
        \caption{\label{fig:overall_design} Overall design of the software system.  Modified from \citet{creaner2016thesis}}
      \end{figure}

The overall design of the software system is outlined in Figure \ref{fig:overall_design}.  The system consists of three major software components and three major data layers.

The software components are as follows:

The \textbf{Parameterisation} software, discussed in brief in Subsection \ref{parameterisation} and expanded upon in \citet{grid_system_paper}, which creates parameter files which define the input data for the other components based on user arguments.

The \textbf{API}, discussed in detail in Subsection \ref{API}, which extracts data from the source catalogue into the local catalogue format.

And the \textbf{Pipeline}, discussed in detail in Subsection \ref{Pipeline} which uses the local catalogue data to identify optimised pointings for each of a set of targets supplied by the user through the parameterisation software and create output catalogue files.

The three main data abstraction layers are as follows:

The \textbf{Source} data is the data generated by previous observers.  In the case of this project, the source data used was SDSS, however as illustrated with the dashed lines in Figure \ref{fig:overall_design}, in principle this system can accommodate any source of data that has the required information about CRS.  The specifics of the SDSS data requirements are outlined in Subsection \ref{SDSS}.

The \textbf{Local} data is data extracted from the Source format for use in this project.  This data can be much reduced in volume compared with the data from the source by eliminating any data redundant to the system.  The Local format is discussed in Subsection \ref{Local}.

The \textbf{Output} data is the results of the project and forms the output catalogues.  This data can be of any format useful to the users of the outputs of this system.  This data is discussed in Subsection \ref{Output}.

\subsection{Parameterisation}
\label{parameterisation}
The Parameterisation software is designed to provide flexible input to the API and Pipeline programs.   Outlined here are the principles of the system and how it impacts the design of the rest of the software used here.

The Parameterisation software is designed to provide flexible input to
the API and Pipeline programs. Provided here is an description of the
principles of parameterisation software design. Specific characteristics
of the parameterisation system, in particular its associated with its
operation upon the SDSS catalogue structure within a Grid HPC system
provided by NGI, are describe in \citet{grid_system_paper}.

The core requirement of the parameterisation system is that it prevent the API and the pipeline from being tied to the structure of a given source catalogue.  This is achieved by having the parameterisation system identify the paths to the input files for a given set of parameters and write those to a parameter file as shown in Subsection \ref{parameter_files}.  These files are then fed into the API and Pipeline, each of which uses the paths specified in the parameter file to find the data it needs.  

In the case of the SDSS system used here, the paths to these files are  based on the SDSS field identifiers (\texttt{run}, \texttt{rerun}, \texttt{camcol} and \texttt{field} as defined in the SDSS Glossary \citep{SDSSGlossary}) which can be used to generate file paths and names in the SDSS directory structure as discussed in Subsection \ref{SDSS}.  This list of identifiers is stored, along with the target data, in a CSV file, structured as shown in Subsection \ref{CSV}.

      \begin{figure}[!htb]
        \center{\includegraphics[width=0.47\textwidth]
        {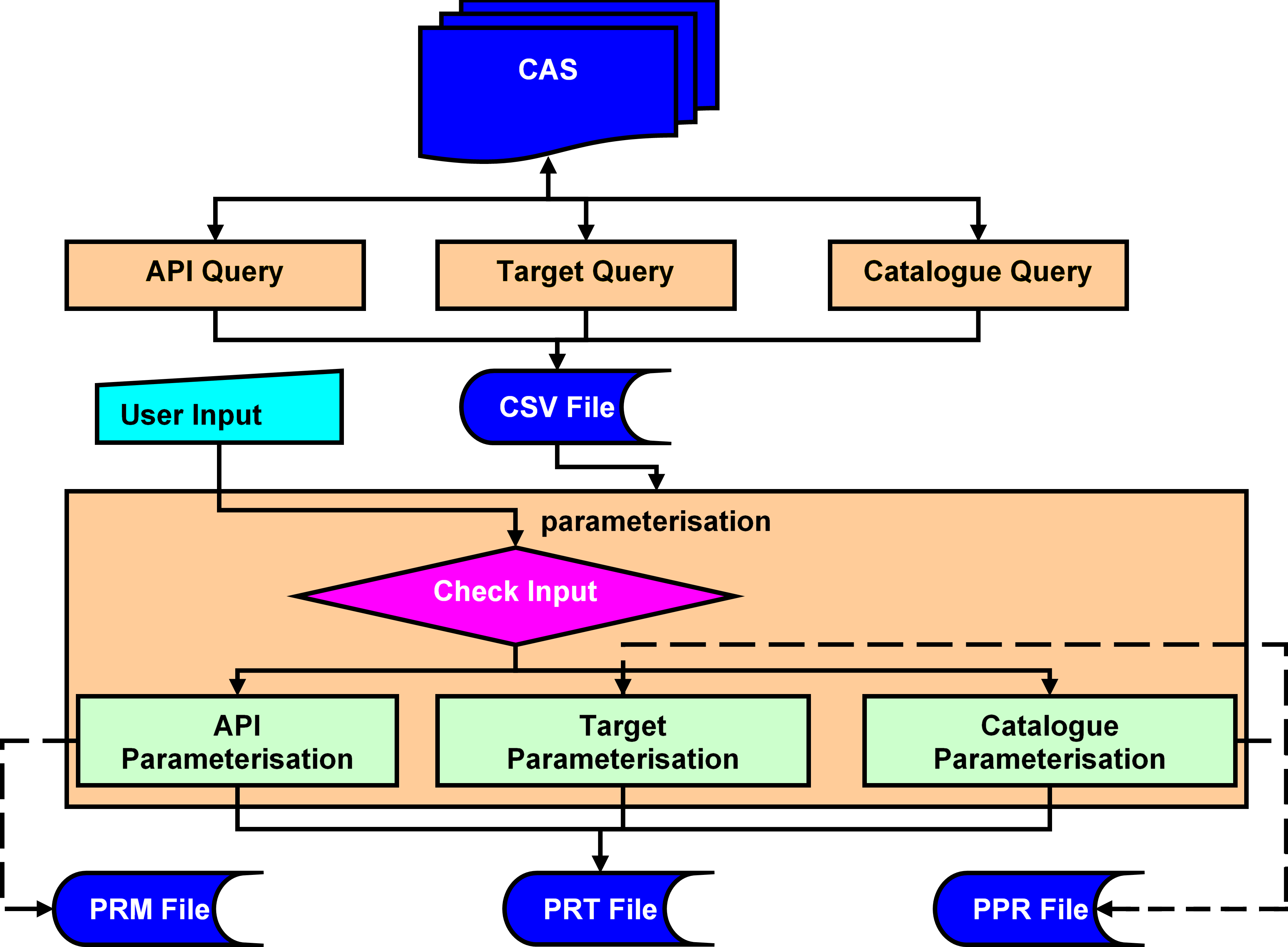}}
        \caption{\label{fig:parameterisation} Parameterisation modes are selected based on the inputs from the user and the nature of the CSV file returned from the CAS.}
      \end{figure}

To generate these field identifiers, SQL Queries were submitted to the SDSS Catalogue Access Server (CAS): an online database which would allow flexible queries to the SDSS structure, but which was not suited to connection to the grid system for frequent access \citep{SDSSCAS}.   Three distinct patterns of query were used for the different use cases of \texttt{parameterisation} as outlined in Figure \ref{fig:parameterisation}.

In \textbf{API} mode the query to the CAS simply selects all fields in the catalogue and returns their field identifiers.

For the Pipeline in \textbf{Target List} mode, a target list consisting of the positions (RA and Dec) of a set of specific targets (e.g. quasars) is submitted through an SQL script to the CAS together with FoV size.  For each target in the list, the script identifies which fields are within the size of the FoV of that target.  The CAS returns a list of field identifiers and the RA and Dec of the targets they refer to in a CSV file as shown in Subsection \ref{CSV}.

For the Pipeline in \textbf{Catalogue Traversal} mode, a list of target fields, consisting of the field identifiers for all fields in the catalogue is submitted to the CAS instead of a target list.  The SQL script identifies, for each target field, the neighbouring fields which have at least one object within the size of the FoV of any point in the target field.  The set of identifiers for the target field are recorded with those of the neighbouring fields in a CSV file as shown in Subsection \ref{CSV}.

In addition to the field identifiers, for the pipeline process, it was necessary to specify a list of targets and the files associated with each of those to fulfil steps 1-3 of the Locus Algorithm as shown in Subsection \ref{requirements}.  As a result, two formats of parameter files were generated as shown in Subsection \ref{parameter_files}, \texttt{PRM} and \texttt{PPR}.  These formats were used in the API and the Pipeline respectively.  As discussed in \citet{grid_system_paper}, the grid management scripts were only able to parse ASCII text files, and as a result, the parameterisation system was modified to also generate text files (labelled \texttt{PRT}) which contained the corresponding paths to the files in the appropriate input catalogue.

The parameterisation software takes user arguments from the command line to specify the number of work units to assign to a given instance of the program as specified in a single parameter file.  Work units are the minimum sized element of the data that can be processed at a time: for the API, the work unit is a single input file from the source catalogue.  For the Pipeline, the inputs are a target or a target field. 

In addition, to operate the parameterisation software in pipeline mode, the user must specify the Observational Parameters FoV size, $\Delta$mag\textsubscript{max}, $\Delta$col\textsubscript{max} and resolution.  These parameters are included in a PPR file as discussed in Subsection \ref{parameter_files}.

\subsection{API}
\label{API}
The API is designed to extract data from the Source catalogue and create the Local Catalogue files used in the pipeline.  As discussed in Subsection \ref{Local}, the Local Catalogue files are designed to only include the minimal set of information about CRS required for the Locus Algorithm.  Thus means that these files are minimised in size, which reduces the amount of data transfer during the pipeline.  

      \begin{figure}[!htb]
        \center{\includegraphics[width=0.3\textwidth]
        {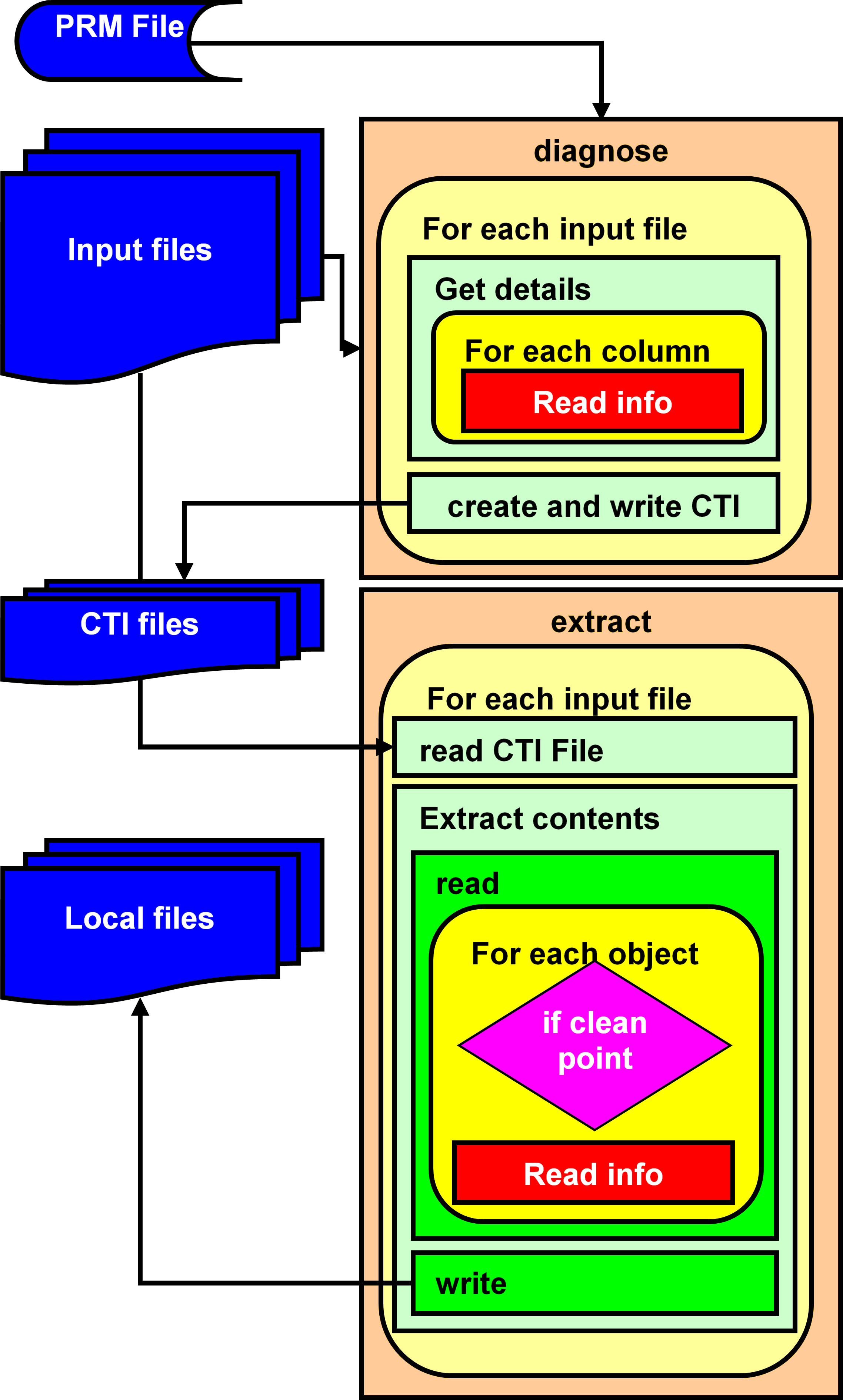}}
        \caption{\label{fig:api} Structure of the SDSS Data ingestion API.  Modified from \citet{creaner2016thesis}}
      \end{figure}

When using SDSS as the source catalogue, the input data is in the form of \texttt{tsObj*.fit} files.  As detailed in Subsection \ref{SDSS}, these files contain many columns of data which are not needed for the algorithm, and many rows which do not pertain to objects that would ever be a suitable CRS.  As a result the API was designed to Select only those rows which relate to suitable objects and project only those columns that were necessary.

As illustrated in Figure \ref{fig:api}, two programs, \texttt{diagnose} and \texttt{extract}, together form the API software system.  These programs use the PRM parameter file defined in Subsection \ref{parameter_files} as an input to access the \texttt{tsObj*.fit} files.  The \texttt{std.out} from \texttt{diagnose} and extract can be verbose as they show the files being read into the program, the selection of fields and the data to be output to the Local Catalogue.  In practice, \texttt{std.out} is typically discarded to \texttt{/dev/null}. 

The \texttt{diagnose} program, accesses and analyses the data contained within the input files, and identifies the columns in the data table.  The column names together with information regarding their structure (\texttt{typecode}, \texttt{repeat} and \texttt{width}) are then stored in a CTI file as defined in Subsection \ref{CTI}. From this information, the key columns can be identified and accessed in \texttt{extract}.

Next, the \texttt{extract} program carries out the substantive work of the API.  Taking the columns identified by \texttt{diagnose}, it identifies the columns containing only the relevant data: Right Ascension, Declination and Magnitude (itself an array of 5 double values).  It then applies the SDSS clean sample of stars algorithm as defined in \citet{SDSScleansample} to exclude entries which are not primary entries for stars.  The data for the remaining entries in those three relevant columns for this project are then copied to a file which forms part of the Local Catalogue.  

A potential refactoring of the code for the API would merge these two programs into one, and in doing so obviate the need for an intermediate file.  This would require substantial reworking of the software including the individual instance software presented here and some modifications grid management scripts as discussed in \citet{grid_system_paper}.

\subsection{Pipeline}
\label{Pipeline}

The Data Pipeline applies the Locus Algorithm as defined in \citet{locuspaper} to a set of targets to produce output files which form the Output Catalogue.  As discussed in \citet{grid_system_paper}, it operates in two modes: Target List and Catalogue Traversal.  These modes differ in how the parameterisation of their targets is handled, but the core operation as illustrated in Figure \ref{fig:pipeline} remains the same in both modes.

Target list mode is used when a set of one or more targets are selected in advance by the user and are submitted to \texttt{parameterisation} to be built into a PPR file.  In this mode, each target is treated independently and separate sets of reference files are used to create the CZ for each target as discussed in subsection \ref{parameter_files}.  This mode was used to produce the Quasar Catalogue as discussed in \citet{quasarpaper}.

Catalogue traversal mode uses an existing catalogue or subset of a catalogue to produce the target list.  In this mode, each file in the catalogue contains a set of targets, and a list of target files from the catalogue is submitted to \texttt{parameterisation} from which a PPR file is to be built.  A single set of reference files is used to create the CZ for all targets in the same target file. This target list is then submitted to the pipeline in a similar way to the Target list mode discussed above.  This mode was used to produce the Exoplanet Catalogue as demonstrated by \citet{ZenodoXOPCatalogue}

\subsubsection{Software Implementation of the Locus Algorithm}
\label{implementation}
\begin{figure}[!htb]
        \center{\includegraphics[width=0.47\textwidth]
        {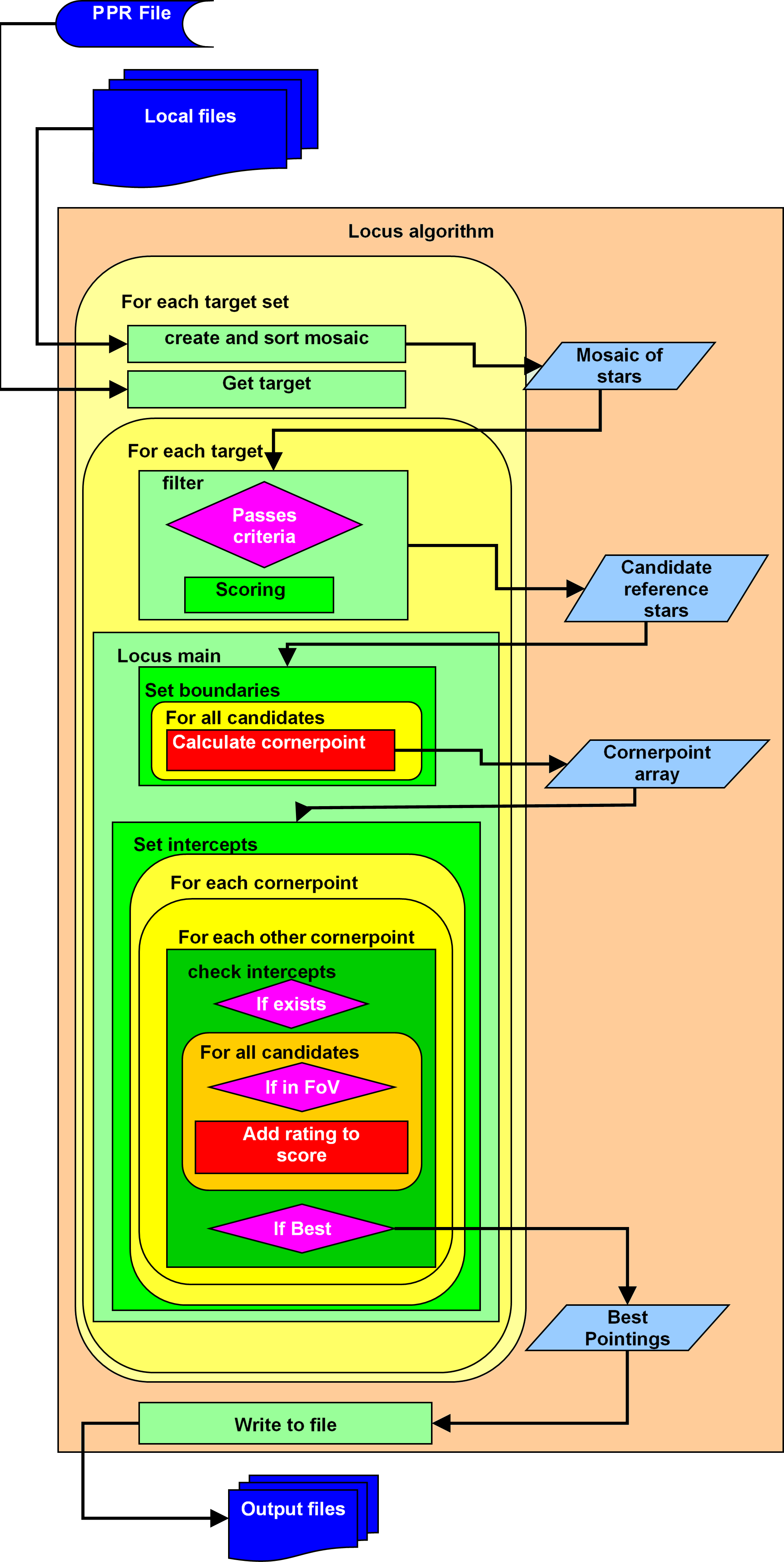}}
        \caption{\label{fig:pipeline} Structure of the Main Data Pipeline.  Modified from \citet{creaner2016thesis}}
      \end{figure}

In this subsection, the step-by step process of how the Pipeline implements the Locus Algorithm is shown.  Here, the numbered steps shown in Subsection \ref{requirements} are referred to at each stage.  Figure \ref{fig:pipeline} illustrates the workflow of this process.

The \texttt{locus\textunderscore{}algorithm} program works as a series of nested loops, illustrated in Figure \ref{fig:pipeline} with darker shades of yellow for inner loops. The outer loops of this system are controlled by data in the \texttt{PPR} file, described in Subsection \ref{parameter_files}.  The inner loops are part of the steps of the algorithm itself.

The outermost loop iterates through target sets.  A target set consists of one or more targets, listed together with the paths to the files in which the data for any stars which could be included in a FoV with any of the targets in the target are stored.  These files are opened and the data read into an array of structures in memory called a \textit{mosaic}.  This mosaic is kept in memory during the processing of each target in the target set, then discarded.  The target set is also read into memory from the \texttt{PPR} file.

The next loop iterates over all targets in the target set.  The target is identified from the \texttt{PPR} file by its RA/Dec coordinates, and its \textit{ugriz} magnitudes, fulfilling \textbf{Step 1} of the Locus Algorithm.  The Observational Parameters are read from the \texttt{PPR} file.  Upper and lower limits for magnitude and colour are calculated from the parameters and the target magnitudes, as per \textbf{Step 2}.  The CZ around the target is defined as a set of upper and lower limits for RA and Dec by reference to the FoV Size and the RA/Dec Coordinates of the target, allowing for a correction factor in RA as described in \citet{locuspaper}, thus meeting the requirements of \textbf{Step 3}.

A function called \texttt{filter} is then called, which compares the data in the mosaic with the limits of the CZ and the observational parameters.  If an object in the the mosaic is within all of the limits set in steps 2 and 3, its data is copied into a new array containing candidate reference stars, completing \textbf{Step 4}.  For each CRS, a \textit{rating} between 0 and 1, which indicates how closely the colour of the CRS matches that of the target is calculated by a function called \texttt{scoring\textunderscore{}mechanism}, as required for \textbf{Step 5}.

The next step, contained in the function \texttt{locus\textunderscore{}main} is to calculate the Loci around each CRS.  As shown in figure \ref{fig:locus} and explained in \citet{locuspaper}, the loci can be defined by their corner-points (CP) in RA and Dec and a direction in which to describe a line each of constant RA and Dec.  The function \texttt{set\textunderscore{}boundaries} defines the corner points by iterating though the CRS and defining the CP by reference to the FoV size and the RA/Dec coordinates of each CRS as defined in \textbf{Step 6} of the Locus Algorithm in \citet{locuspaper}.  The direction of the lines of constant RA/Dec are stored as a binary switch with 0 indicating the negative direction and 1 indicating a positive direction.

\textbf{Step 7} is accomplished by the function \texttt{set\textunderscore{}intercepts}.  It iterates over each CP, and within that loop, iterates over each other CP, combining the RA of the first CP with the Dec of the second CP to define a potential PoI.  The function \texttt{check\textunderscore{}intercepts} then uses the direction variables to determine whether the lines of constant RA and Dec actually cross.  If so, a PoI exists at that combination of RA and Dec.

The software then iterates over each CRS and checks whether it could be included in a FoV centred on that PoI.  If so, its rating is added to the score for that PoI.  This completes \textbf{Step 8}.

A pair of structured variables containing the RA, Dec and Score of the pointings \texttt{current\textunderscore{}pointing} and \texttt{best\textunderscore{}pointing} (initialised to [0.0,0.0,0.0]) are used to track which PoI produces the best score.  If a new PoI produces a better score than the previous best pointing, then the new PoI replaces the previous one as \texttt{best\textunderscore{}pointing}.  Once all combinations of CPs have been tested, the final \texttt{best\textunderscore{}pointing} is copied to an array of pointings, which rounds out the Algorithm at \textbf{Step 9}.

The program then iterates to the next target in the target set.  If the target set is completed, it moves to the next target set and creates a new mosaic.  If all target sets have been processed, the array of optimum pointings is written to a \texttt{FITS} file and the program closes.  This file can be made available as part of the Output Catalogue.

\section{Data Structures and Management}
\label{Data}

The system for this project was built with flexibility of inputs and data abstraction as key design ambitions.  First, in Subsection \ref{Formats} the different designs of data storage formats used in the project are discussed.  Then,  Subsection \ref{layers} considers how those were accommodated by means of separating the data into three layers of abstraction, each of which filled a different role in the project.   The physical and logical layout of the storage media used to store the data for this project are discussed in \citet{grid_system_paper}, as they pertain more to the specific grid implementation of the system than an overarching design principle.

\subsection{Data Formats}
\label{Formats}
A variety of data formats were used throughout the course of the project.  These formats had some impact on the design of the project.  Some of these formats (FITS [\ref{FITS}], CSV [\ref{CSV}] and JDL [\ref{JDL}])  were pre-existing and developed outside the project, others (PRM, PPR, PRT [\ref{parameter_files}] and CTI [\ref{CTI}]) were developed within the project.  For the former, the type is introduced briefly, but the Subsections below explain how it was used in this project.  For the latter, the motivation for the datatype's developement is discussed, and details are given of its internal structure as well as its use within the project.

\subsubsection{FITS}
\label{FITS}

\begin{figure}[!htb]
        \center{\includegraphics[width=0.47\textwidth]
        {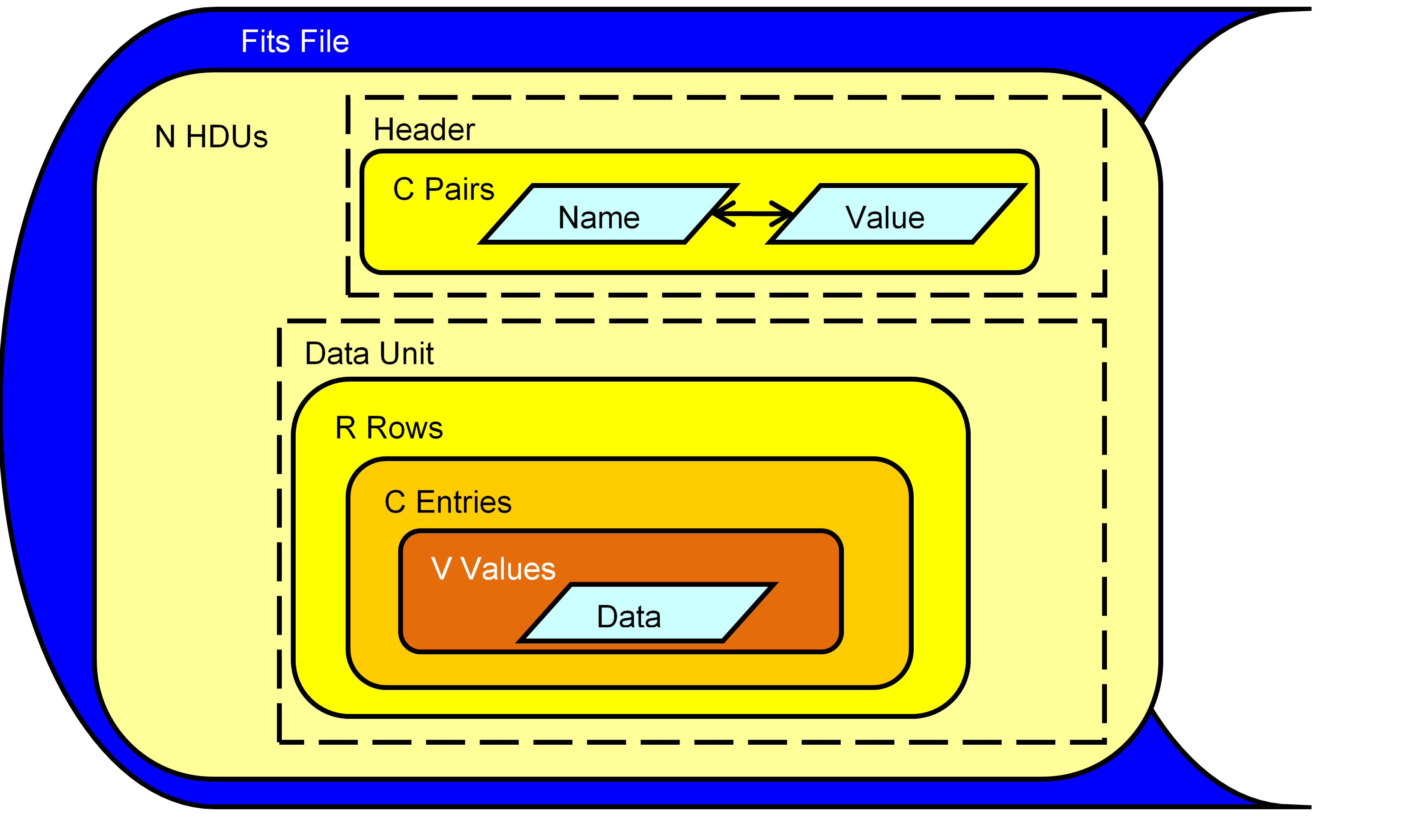}}
        \caption{\label{fig:fits} General structure of a \texttt{FITS} file.  Modified from \citet{creaner2016thesis}}
      \end{figure}

FITS stands for Flexible Image Transport System.  FITS files, (extension \texttt{.fit}) are a common standard in the for the transfer of astronomical data (including images) \citep{pence1999cfitsio}. FITS files made of a series of HDUs (Header Data Units) which contain a header and some data \citep{pence2010definition}. Headers  are a sequence of lines of the form \texttt{keyword=value} that describe the structure of the data and the data types within it.  Headers may also include additional information about the instruments used or the history of the data.  Data follows on from the header, and is structured as it is described in the header \citep{cfitsiouserguide}.  Data may consist of images, data tables, or data cubes based on the requirements of the project \citep{pence2010definition}.  In general, the structure follows the layout described in Figure \ref{fig:fits}.

All of the various uses of the FITS format in this project are built as a single HDU which contains a data table.  FITS files were used as the source data for the project, with data from the SDSS DAS as illuestrated in Subsection \ref{SDSS}, the intermediate Local Catalogue, used to abstract source data for use in the Pipeline as discussed in Subsection \ref{Local} and outputs for the project were intitially generated in FITS format as discussed in Subsection \ref{Output}.

\subsubsection{CSV}
\label{CSV}

Comma-Separated Variable (CSV) files are plain-text (e.g. ASCII or UTF-8) files in which data tables are stored by deliminating records by newline characters and fields within that record by some delimiter, usually (but not always) a comma \citep{shafranovich2005common}.  As a result, this format can be accessed and read by most common text editors, and parsed by low-level systems such as the BASH shell script interpreter.  It is common for the first line of these files to contain a header consisting of strings which are used to label the columns of the data.

\begin{table}[htb!]
\centering
\begin{tabular}{c c }
\hline
   \textbf{CSV Use-Case} & \textbf{Fields} \\ \hline \hline
\textbf{SDSS} \textbf{API} & \texttt{run}, \texttt{rerun}, \texttt{camcol}, \texttt{field}     \\ \hline
\makecell{\textbf{Target} \textbf{List}\\\textbf{Pipeline}} &  \makecell{ \texttt{ObjectID}, \texttt{FieldNum},\\ \textit{reference}[\texttt{run}, \texttt{rerun}, \texttt{camcol}, \texttt{field}],\\ \textit{target}[\texttt{RA}, \texttt{Dec}, \textit{magnitudes}\{\texttt{u}, \texttt{g}, \texttt{r}, \texttt{i}, \texttt{z}\}]  }   \\ \hline
\makecell{\textbf{Catalogue}\\ \textbf{Traversal}\\\textbf{Pipeline}} & \makecell{\texttt{MosaicNo}, \texttt{FieldNum},  \\\textit{reference}[\texttt{run}, \texttt{rerun}, \texttt{camcol}, \texttt{field}], \\\textit{target}[\texttt{run}, \texttt{rerun}, \texttt{camcol}, \texttt{field}]} \\ \hline 
\makecell{\textbf{Duplicate}\\\textbf{Output File}} &  \makecell{ \textit{target}[\texttt{RA}, \texttt{Dec}\textit{magnitudes}\{\texttt{u}, \texttt{g}, \texttt{r}, \texttt{i}, \texttt{z}\}],\\ \textit{pointing}[\texttt{RA}, \texttt{Dec}, \texttt{score}]  }   \\ \hline
\end{tabular}

 \caption{Table of the difference Use-Cases for CSV files, together with the fields contained within each record.  Logical groupings of fields are shown in italics, enclosed in square brackets or braces.}
 \label{table:csv} 
\end{table}

In this project, CSV files are produced as output from the CAS \citep{SDSSCAS}, and are used as inputs from there to \texttt{parameterisation}.  As shown in Table \ref{table:csv}, there are three structures of CSV output from the CAS and each of them contains different sets of information.  

For the API, a list of field identifiers is produced to allow the API to iterate through all files in the SDSS catalogue by allowing \texttt{parameterisation} to generate the paths to those files as illustrated in Subsections \ref{parameterisation} and \ref{SDSS}.  

In the Target List mode, a set of targets is passed to the CAS, and for each a set of reference fields is generated.  Thus the system returns target data and the corresponding field identifiers.  A limitation of the CSV format which means it is unsuitable for hierarchical or relational data means that the target information is repeated on each row listing the field identifiers required to make a mosaic around that target.  This is then parsed by \texttt{parameterisation} to allow the pipeline to find the paths to the files containing reference stars and to ingest the information regarding the target.

In the Catalogue Traversal mode, a set of target field identifiers is passed to the CAS.  The CAS identifies reference fields which can be used to generate mosaics around any targets in the target fields.  As before, this leads to some repetition of data in the CSV file that is returned to be parsed by \texttt{parameterisation} in a similar way, again allowing the Pipeline to produce pointing.

Finally, CSV files are generated containing duplicate information to that contained in the Output files generated by the Pipeline to allow for easy access to the data for end users.  The structure of this data is also shown on Table \ref{table:csv}.

\subsubsection{JDL}
\label{JDL}

\begin{figure}[!htb]
        \center{\includegraphics[width=0.47\textwidth]
        {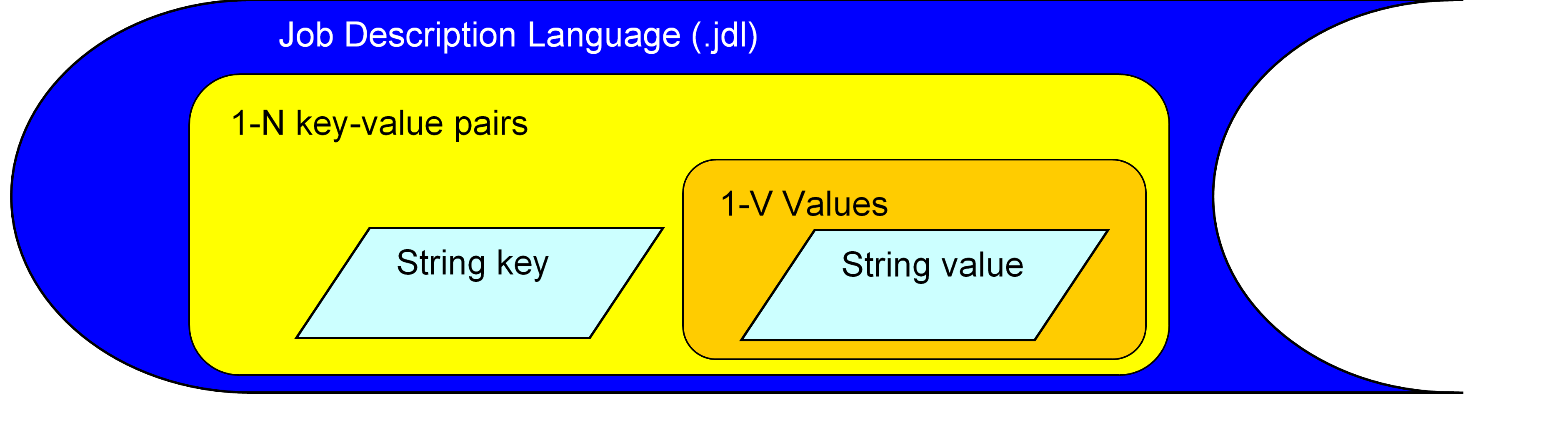}}
        \caption{\label{fig:jdl} General structure of a \texttt{JDL} file.  Modified from \citet{creaner2016thesis}}
      \end{figure}

Job Description Language (JDL) files are used to pass instructions to the \texttt{glite} grid middleware system \citep{glite} used in this project as part of the NGI HPC solution used \citep{coghlan2005grid}.  The attributes of grid jobs are specified as \texttt{key=value} pairs.  The keys are drawn from a list of grid job attributes defined by the grid managers, while the values are strings which specify the attribute for that specific job. (e.g. the executable to be run might be specified as \texttt{executable = "call\textunderscore{}api.sh"}).  This format is illustrated in Figure \ref{fig:jdl}.

In the course of this project, many JDL files were automatically generated by a suite of job management software.  This software, written as BASH shell scripts is described in detail in \citet{grid_system_paper}.

\subsubsection{Parameter Files (PRM, PPR \& PRT)}
\label{parameter_files}
Parameter files are used in this project to provide the API and the Pipeline with the input data used in each without requiring either system to interpret the data storage structure the input data is held in.  As shown in Subsection \ref{parameterisation}, this is achieved by having \texttt{parameterisation} parse catalogue specific terminology such as the SDSS field identifiers into paths to the data as stored in the system.  The core programs can thus read the path from the parameter file and then open the data file at that location for processing.  PRM and PPR files are binary files, allowing for a variety of datatypes to be stored in a compact manner.  PRT files are ASCII-text files which can only contain text (or text representations of other data).

\begin{figure}[!htb]
        \center{\includegraphics[width=0.47\textwidth]
        {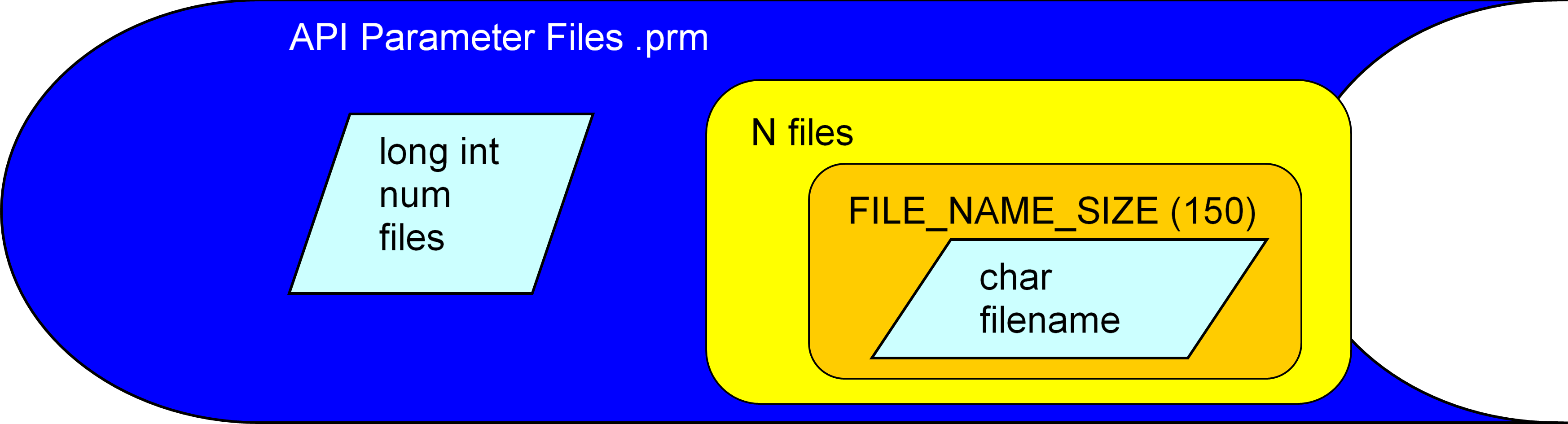}}
        \caption{\label{fig:prm_files}Schematic of the structure of a PRM file.  The file consists of a number of files to look for, and the paths to each of those files}
      \end{figure}

\textbf{API Parameter files} (PRM) are structured with a \texttt{long int}-datatype number as a loop-control variable, indicating the number of files to be processed by the API, and then a series of 150-\texttt{char} strings for the file paths, as illustrated in Figure \ref{fig:prm_files}.  This allows for quick- and consistent parsing of the data in the PRM file and thus allows the system to move on to the reading of the data from the FITS files quickly. 

\begin{figure}[!htb]
        \center{\includegraphics[width=0.47\textwidth]
        {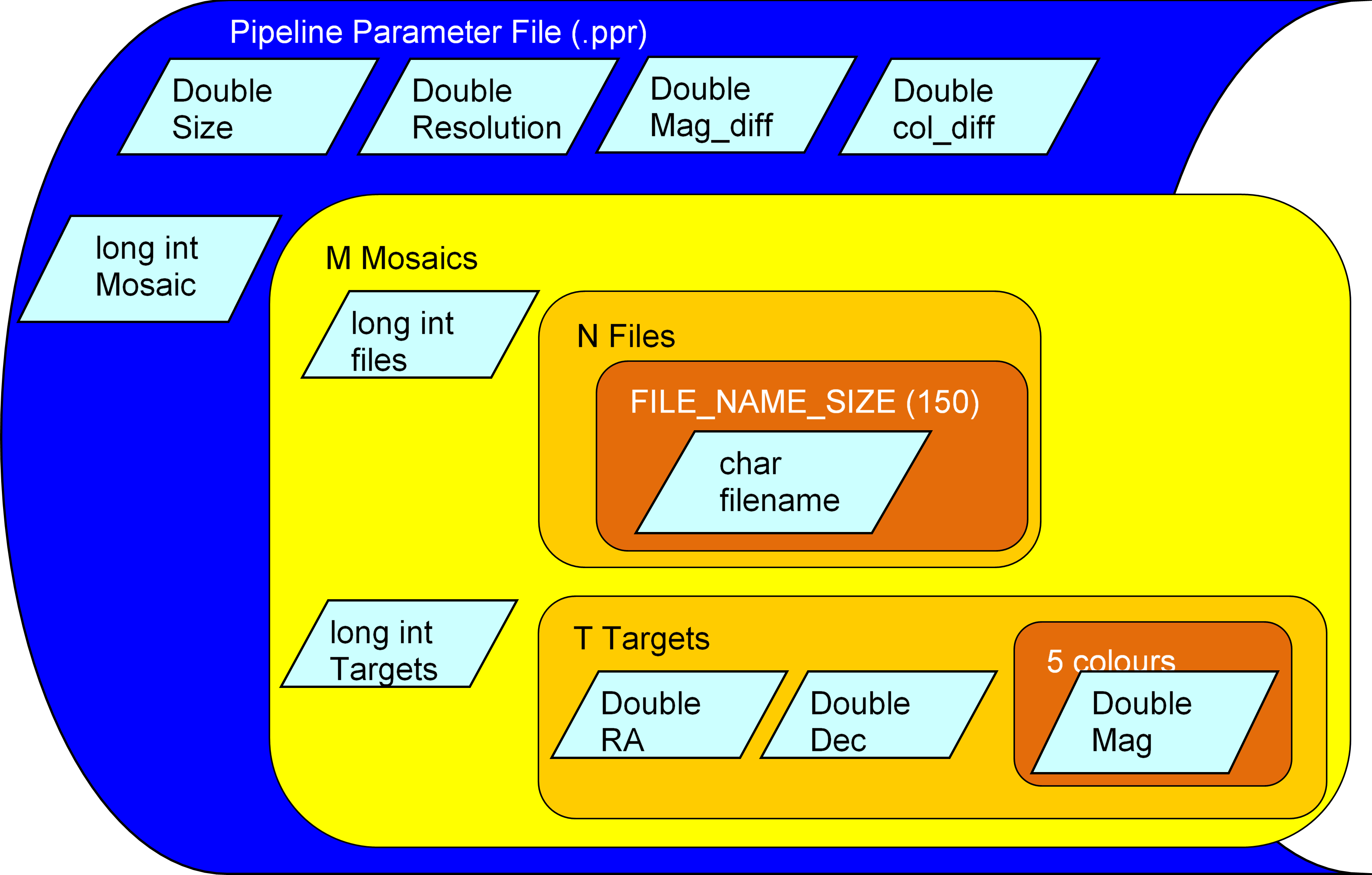}}
        \caption{\label{fig:ppr_files}Schematic of the structure of a PPR file.  The file consists of four observational parameters, a number indicating the number of grouped sets of targets, a list of the paths to the fields near each group, and a list of the position and magnitudes for each of the targets in each of the groups}
      \end{figure}

\textbf{Pipeline Parameter files} are structured with data that iterates over several loops as illustrated in Figure \ref{fig:ppr_files}.  First, the observational parameters which are common to all pointings within a use of the project, FoV Size, Resolution, $\Delta$mag\textsubscript{max} and $\Delta$col\textsubscript{max} are provided, all stored as \texttt{double}-datatype floating point numbers.  Next, a number indicating how many target sets and the mosaics required for those targets is provided as a loop-control variable.  The outermost loop then iterates that many times.  Within that loop, there are two blocks, the mosaic and the target.  The mosaic blocks begin by first giving a loop-control variable, the \texttt{long int} number of files required to create the mosaic, then, as with the PRM files above, iterating through a series of 150-\texttt{char} strings for the file paths.  The target block begins with a loop-control variable, the \texttt{long int} number of targets that form that target set.  (Note: when the Pipeline is used in Target List mode as discussed in Subsection \ref{Pipeline}, this number of targets is 1.)  The target block then iterates over the targts, giving the RA, Dec and Magnitudes (\textit{ugriz}, in the case of SDSS) of the targets, all formatted as \texttt{double}-precision numbers.

Finally, PRT files were developed because the grid management scripts, written in the BASH shell scripting language, were not able to easily parse the paths contained in the binary format PRM and PPR files.  This meant that when data had to be moved between different storage elements as dicussed in \citet{grid_system_paper}, it was necessary to provide these paths in a more readable format.  PRT files are plain-text files, generated by \texttt{parameterisation} at the same time as the PRM and PPR files. These contain the same paths stored in the binary files, one per line of the PRT file, but none of the loop-control variables, observational parameters or target data.

\subsubsection{CTI}
\label{CTI}
CatalogueInformation files (CTI) are used interally to the API to pass data about the contents of a given input file between \texttt{diagnose} and \texttt{extract} as shown in Subsection \ref{API}. Because the structrure of files in the SDSS catalogue can change between data releases, \texttt{diagnose} intially examines the input file, and extracts a description of that file which is stored within the CTI file.  That description is then passed to \texttt{extract} for further processing.
      \begin{figure}[!htb]
        \center{\includegraphics[width=0.47\textwidth]
        {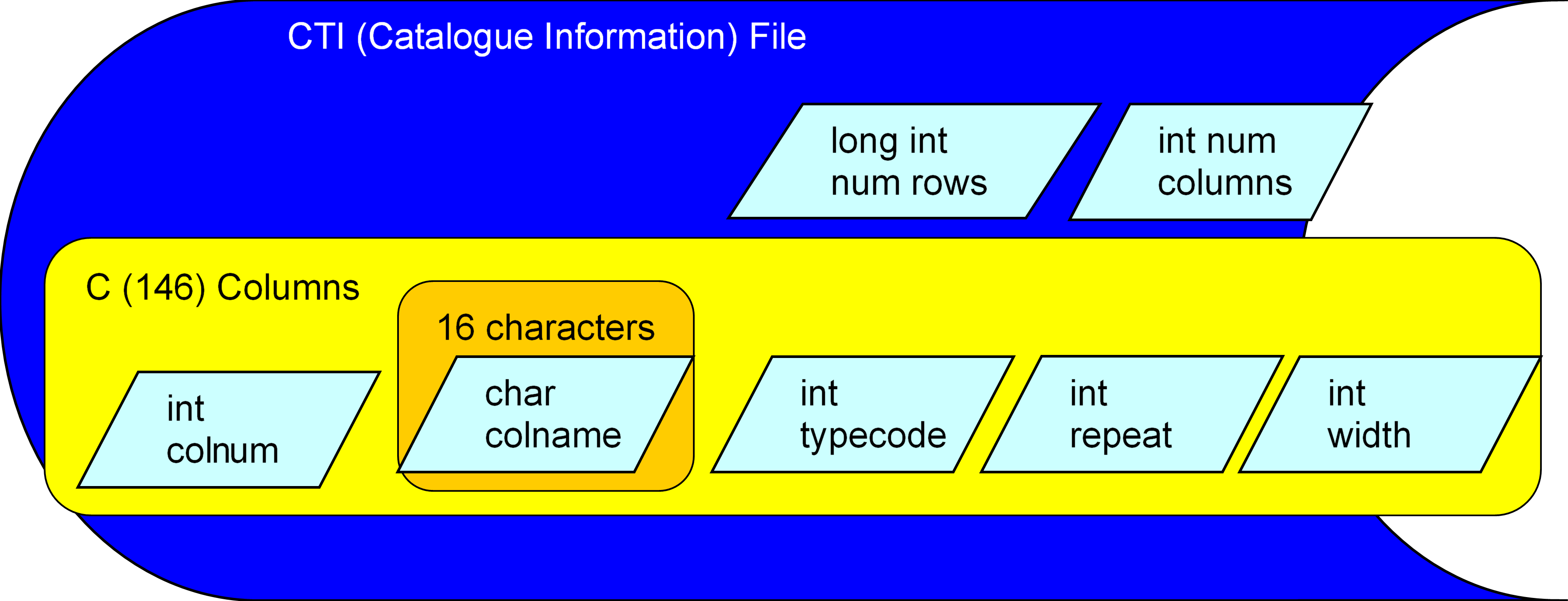}}
        \caption{\label{fig:cti}Schematic of the structure of a CTI file.  }
      \end{figure}

The CTI file consists of a \texttt{long int} for the number of rows and an \texttt{int} for the columns within that file.  It then iterates through the columns, and for each gives an \texttt{int} for the coumn number, a 16-\texttt{char} string for the column name, and three \texttt{int} values for \texttt{typecode}, \texttt{repeat} and \texttt{width}, internal descriptions of the structure of the data in ech column.

A proposed refactoring of the system would merge \texttt{diagnose} and \texttt{extract} into a single program, and potentially render the CTI datatype redundant as discussed in Subsection \ref{API}.

\subsection{Data Layers}
\label{layers}
As illustrated in Figure \ref{fig:overall_design}, in this project, a minimal set of data required for the algorithm is extracted from Source Catalogues (such as SDSS) into the Local Catalogue.  In doing so, the volume of data is reduced considerably, allowing for repeat access to the same data over network connections much more rapidly than would be possible if the full source catalogue was used.  The use of SDSS as a source catalogue, and the design implications this had for the project is considered in Subsection \ref{SDSS}.

This local catalogue data was then used as inputs to the pipeline which calculated sets of optimal pointings.  The Local Catalogue format is designed to be used with any source catalogue with minimal changes, but certain design elements are constrained by the data in the source catalogue and thus the details of the Local Catalogue as used in Subsection \ref{Local} are somewhat influenced by SDSS.

Finally, the outputs of this project are sets of targets and pointings, and are largely organised based on the processing jobs, each consisting of multiple work units as discussed in Subsetion \ref{parameterisation}.  The format of these outputs are largely unconstrained by the formats of the inputs.  The choice of data and file structures for the output is outlined in Subsection \ref{Output}.

\subsubsection{SDSS}
\label{SDSS}
Extracting data from the source catalogue is necessarily different for each source catalogue.  In the case of SDSS, the full collection of Object Catalogue files, labelled \texttt{tsObj*.fit}, were downloaded from the Data Archive Server (DAS) using the \texttt{wget} UNIX command.  

      \begin{figure}[!hbt]
        \center{\includegraphics[width=0.47\textwidth]
        {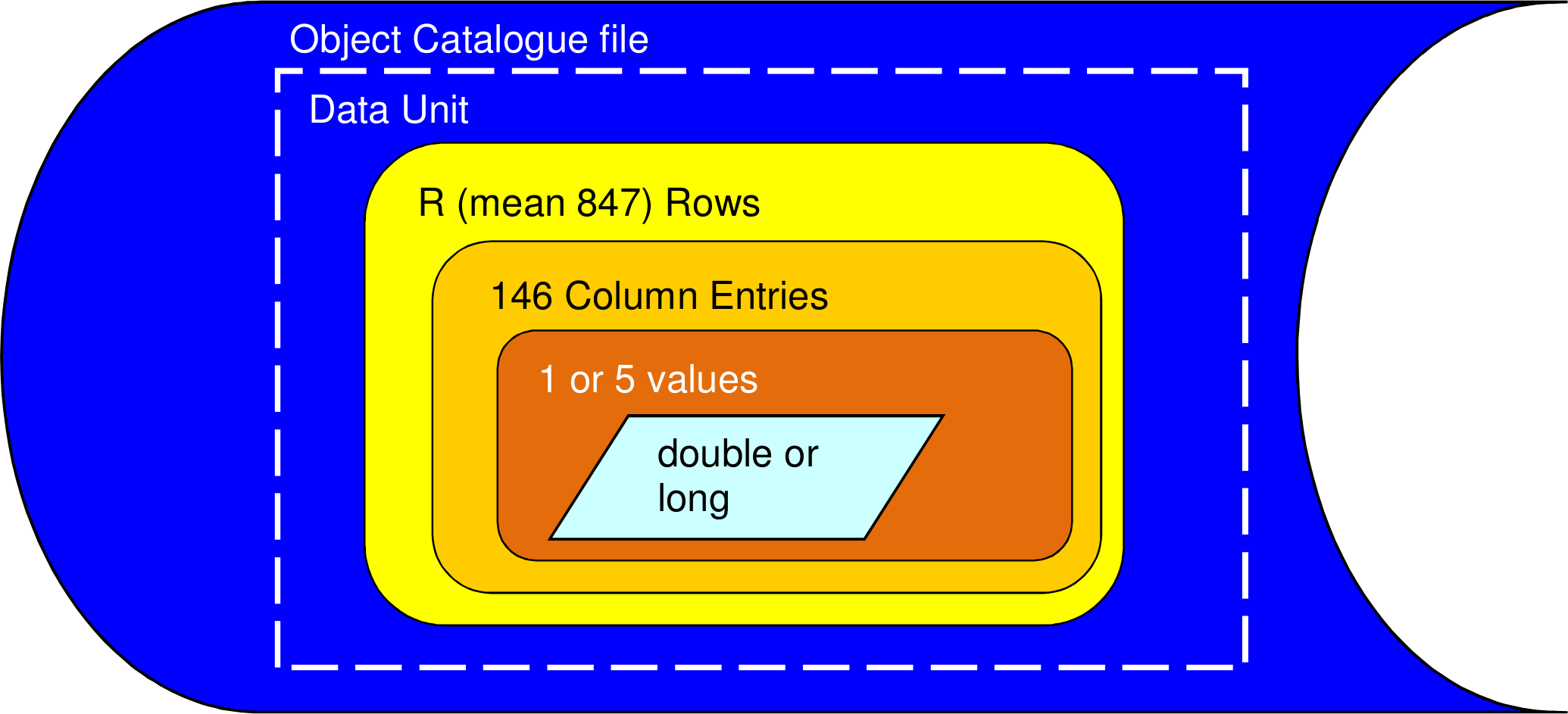}}
        \caption{\label{fig:tsObj} Internal structure of a \texttt{tsObj*.fit} file.  Modified from \citet{creaner2016thesis}}
      \end{figure} 

The Source files contain records of data about every detection made using the SDSS telescope and camera.  These files are structured as FITS files as discussed in Subsection \ref{FITS} consisting of a single Header Data Unit (HDU) containing a data table.  This table consists of a variable number of rows (mean rows 847) and 146 data columns.  Those columns may be a single value of an atomic data type (\texttt{double} or \texttt{long int}) or an array of 5 such values (one for each of the SDSS \textit{ugriz} filters).  This structure is illustrated in \ref{fig:tsObj}.

The files in the DAS were stored in a directory structure and in filenames that were built from the SDSS field identifiers \texttt{run}, \texttt{rerun}, \texttt{camcol} and \texttt{field} \citep{SDSSDAS}.  This allowed the correct file for a given position on the sky to be identified without needing to open the files and parse the data within them using SQL queries to the CAS as discussed in Subsection \ref{parameterisation} \citep{SDSSCAS}.

Using the following convention:
\begin{center}
\textcolor{OliveGreen}{\texttt{run}} = \textcolor{OliveGreen}{\texttt{r}}

\textcolor{Maroon}{\texttt{rerun}} = \textcolor{Maroon}{\texttt{R}}

\textcolor{MidnightBlue}{\texttt{Camcol}} = \textcolor{MidnightBlue}{\texttt{C}}

\textcolor{Orange}{\texttt{field}} = \textcolor{Orange}{\texttt{F}} 

And using repeated letters to indicate padding with leading zeroes as needed
\end{center}

The directory structure is thus given by:
\begin{center}
\textbf{\texttt{\textcolor{Maroon}{R}/\textcolor{OliveGreen}{r}/calibChunks/\textcolor{MidnightBlue}{C}/}}

e.g. \textbf{\texttt{\textcolor{Maroon}{1458}/\textcolor{OliveGreen}{40}/calibChunks/\textcolor{MidnightBlue}{4}/}}
\end{center}

Similarly, the filename is defined as:
\begin{center}
\textbf{\texttt{tsObj-\textcolor{Maroon}{RRRRRR}-\textcolor{MidnightBlue}{C}-\textcolor{OliveGreen}{r}-\textcolor{Orange}{FFFF}.fit}}

e.g. \textbf{\texttt{tsObj-\textcolor{Maroon}{001458}-\textcolor{MidnightBlue}{4}-\textcolor{OliveGreen}{40}-\textcolor{Orange}{0352}.fit}}
\end{center}

Automatic parsing of this structure was needed to access the data in the source catalogue.  This was provided through the \texttt{parameterisation} program as discussed in Subsection \ref{parameterisation}.  

\subsubsection{Local Catalogue}
\label{Local}
The records in the \texttt{tsObj*.fit} files include spurious detections such as Cosmic Rays, non-stellar sources unsuitable for use as reference stars, and secondary detections, where the same object has been detected multiple times (e.g. when several rectangular SDSS \texttt{fields} overlap due to the tiling pattern on the sky).  For each record of a detected object there are many columns containing data such as Stokes Parameters and regarding the shape of non-stellar objects which are not pertinent to this project.  

The local format, by contrast consists of just three columns: the position (RA and Dec) and magnitude.  For data extracted from SDSS, magnitude was stored as an array of five values, one for each of the \textit{ugriz} magnitudes.  In generating the local catalogue, the SDSS Clean Sample of Point Sources algorithm \citep{SDSScleansample} was applied to eliminate any extraneous records as shown in Subsection \ref{API}.

When the data was extracted from the source catalogue, it needed to be stored in a format that could be accessed by the pipeline.  To reduce the number of library dependencies in this project, it was decided to use the FITS format for this task as well.  

Further to this, the pipeline would need to be able to identify which files to use with which targets.  It was decided to retain the SDSS file naming convention and directory structure, with the file extension \texttt{.fit} replaced with \texttt{\textunderscore{}local.fit}, such that SQL queries to the CAS could be used to identify files based on the location of targets as shown in Subsection \ref{parameterisation}.  

Note that the layer of abstraction provided by \texttt{parameterisation} means that subsequent iterations of the system can use different source catalogues with different data structures and local catalogue files can be generated to match.  The pipeline will be able to accommodate this structure without changes, by replacing the modular components of \texttt{parameterisation} with ones which guide the system to the new data.

      \begin{figure}[!htb]
        \center{\includegraphics[width=0.47\textwidth]
        {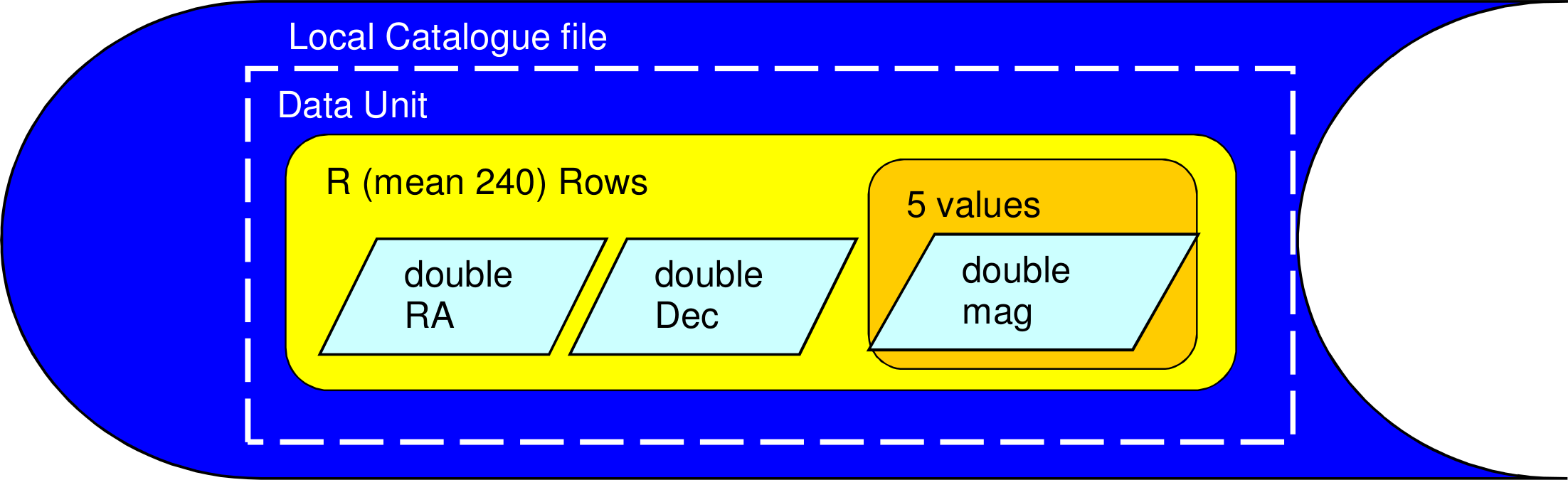}}
        \caption{\label{fig:local_fits} Internal structure of a local catalogue fits file.  Modified from \citet{creaner2016thesis}}
      \end{figure} 
      
As a result, a system was created where each source file had a corresponding local file, but with reduced rows and columns. This produced local FITS files with a single HDU which included one data table of three columns and with a mean number of rows of 240 as illustrated in Figure \ref{fig:local_fits}.

Between these factors, the data volume of the Source catalogue was reduced by a factor of 707 in producing the Local Catalogue as discussed in \citet{grid_metrics_paper}.
\subsubsection{Output Catalogues}
\label{Output}

\begin{figure}[!htb]
        \center{\includegraphics[width=0.47\textwidth]
        {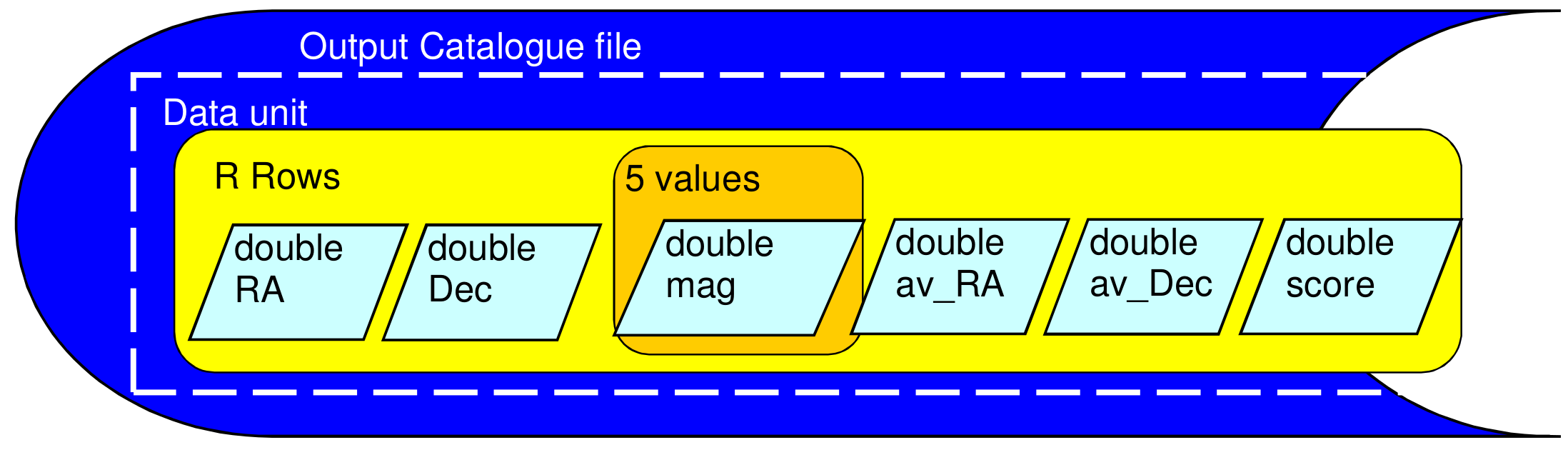}}
        \caption{\label{fig:out_fits} Internal structure of an output catalogue fits file.  Note the labels \texttt{av\textunderscore{}RA} and \texttt{av\textunderscore{}Dec} refer to the pointing.  Modified from \citet{creaner2016thesis}}
      \end{figure}

Output data from the software is a series of records pertaining to the target(s) and the optimum pointings for each target.  This consists of six columns: the RA, Dec and magnitudes (\textit{ugriz} in the case of SDSS) of the Target and the RA, Dec and score for the pointing.  Because output data is generated by jobs consisting of many work units (targets or target sets) as discussed in Subsections \ref{parameterisation} and \ref{Pipeline}, the number of such records varied based on the inputs, and was independent of the SDSS data structure. the file structure is illustrated in \ref{fig:out_fits}.

For consistency in application, FITS was again chosen as the output format for the pipeline.  Output files were named for the job that generated them, and as discussed in \citet{grid_system_paper}, jobs were automatically generated, including the job names.  

In making the catalogues available to the public, the raw FITS files were made available, but it was also decided to make copies of the files available in the more accessible CSV format.  Finally, to aid with database ingestion and querying, the short CSV files were concatenated to produce larger CSV files containing the entire catalogue, however these are unsuitable for processing by systems with low memory capacity.  These catalogues are available at \citep{ZenodoQuasarCatalogue,ZenodoXOPCatalogue}.

\section{Scaling to Catalogues}
\label{Scaling}

Initial tests using this system to identify pointings for single targets demonstrated that the program had a runtime of the order of 1 second on commercial-off-the-shelf hardware with samples of the data downloaded and stored in the local filesystem.  When comparing this against the "357 million unique objects" in the SDSS DR7 Catalogue used at the outset of this project, several scaling issues became apparent \citep{abazajian2009seventh}.

Data scaling would prove difficult for common equipment available at the time: the 4.2TB of the DAS calibrated objects catalogue \citep{SDSSDAS} would not be possible to store at once, and some data staging to removable or network storage would be necessary.

Processing was expected to scale approximately linearly in most circumstances: this lead to predictions of overall run time to process the entire SDSS catalogue which were of the order of 10 years.

It is thus apparent that a high-performance computing solution would be needed to allow for large data storage requirements to be met, and for parallelisation of data processing to overcome data processing.  The solution to this challenge is discussed in \citep{grid_system_paper}.

\section{Conclusions}
\label{Conclusions}

The software system presented here meets the Requirements, observes the Constraints and achieves the Ambitions set out in Section \ref{Considerations} of this paper.

The API, discussed in Subsection \ref{API} reads data in from a source catalogue, SDSS, into a local format which can be used by the rest of the system.  The Locus Algorithm, defined in \citet{locuspaper} provides the solution to identifying potential pointings from within the data and comparing them with one another through the respective scores for each pointing.  This algorithm is implemented step-by-step by the Pipeline as shown in Subsection \ref{Pipeline}.  The targets and their pointings are output in FITS and CSV format as dicussed in Subsection \ref{Output} and presented in \citet{quasarpaper,ZenodoQuasarCatalogue,ZenodoXOPCatalogue}.

The constraints imposed by the source data from SDSS were accommodated by providing layers of data abstraction from the source as discussed in Subsection \ref{layers}.  Programming in the highly portable C language mitigated any constraint required by the UNIX operating system employed at NGI.  The use of grid scripts to ``wrap'' the software presented here within and interact with the \texttt{glite} middleware allowed the rest of the system to be developed without constraint from this design element, and enabled testing to be performed on a non-grid system for easier development and testing cycles.

The ambition to have flexible software was achieved by having allowing the user to specify parameters at runtime.  The software is extensible to allow extra elements to be added without a complete overhaul in most cases.  Modular design enabled separation of roles between different elements of the sytem, allowing for reuse of modules as well as replacement with new versions as required.  Data abstraction means that the same system can be used for a variety of input data types and systems with only minimial changes to existing software.

The software used in this project is available at on Github at \citet{githubrepo}.  A paper describing how this system is implemented using grid computing is available at \citet{grid_system_paper} and the performance metrics of that system are available at \citet{grid_metrics_paper}.

\section*{Acknowledgements}
\textbf{Funding for this work}: This publication has received funding from Higher Education Authority Technological Sector Research Fund and the Institute of Technology, Tallaght, Dublin Continuation Fund (now Tallaght Campus, Technological University Dublin).

\textbf{SDSS Acknowledgement}: This paper makes use of data from the Sloan Digital Sky Survey (SDSS).  Funding for the SDSS and SDSS-II has been provided by the Alfred P. Sloan Foundation, the Participating Institutions, the National Science Foundation, the U.S. Department of Energy, the National Aeronautics and Space Administration, the Japanese Monbukagakusho, the Max Planck Society, and the Higher Education Funding Council for England. The SDSS Web Site is http://www.sdss.org/.

The SDSS is managed by the Astrophysical Research Consortium for the Participating Institutions. The Participating Institutions are the American Museum of Natural History, Astrophysical Institute Potsdam, University of Basel, University of Cambridge, Case Western Reserve University, University of Chicago, Drexel University, Fermilab, the Institute for Advanced Study, the Japan Participation Group, Johns Hopkins University, the Joint Institute for Nuclear Astrophysics, the Kavli Institute for Particle Astrophysics and Cosmology, the Korean Scientist Group, the Chinese Academy of Sciences (LAMOST), Los Alamos National Laboratory, the Max-Planck-Institute for Astronomy (MPIA), the Max-Planck-Institute for Astrophysics (MPA), New Mexico State University, Ohio State University, University of Pittsburgh, University of Portsmouth, Princeton University, the United States Naval Observatory, and the University of Washington.




\bibliographystyle{elsarticle-harv}
\bibliography{software_paper}







\end{document}